\documentclass[%
preprint,
 amsmath,amssymb,
 aps,prl,superscriptaddress
]{revtex4-1}

\usepackage{graphicx}
\usepackage{dcolumn}
\usepackage{bm}
\usepackage[mathlines]{lineno}
\usepackage{floatrow}
\usepackage[bottom]{footmisc}
\begin{document}

\preprint{APS/123-QED}

\title{The $^{7}$Be($\boldsymbol{n,p}$)$^{7}$Li reaction and the Cosmological Lithium Problem: measurement of the cross section in a wide energy range at n\_TOF (CERN)}

\author{L.~Damone} \affiliation{INFN, Sezione di Bari, Italy} \affiliation{Dipartimento di Fisica, Universit\`{a} degli Studi di Bari, Italy}
\author{M.~Barbagallo} \affiliation{INFN, Sezione di Bari, Italy} \affiliation{European Organization for Nuclear Research (CERN), Switzerland}
\author{M.~Mastromarco} \affiliation{INFN, Sezione di Bari, Italy} \affiliation{European Organization for Nuclear Research (CERN), Switzerland}
\author{A.~Mengoni}\thanks{Corresponding author: alberto.mengoni@enea.it} \affiliation{ENEA, Bologna, Italy} \affiliation{INFN, Sezione di Bologna, Italy}
\author{L.~Cosentino} \affiliation{INFN - Laboratori Nazionali del Sud, Catania, Italy}
\author{E.~Maugeri} \affiliation{Paul Scherrer Institut, Villigen PSI, Switzerland}
\author{S.~Heinitz} \affiliation{Paul Scherrer Institut, Villigen PSI, Switzerland}
\author{D.~Schumann} \affiliation{Paul Scherrer Institut, Villigen PSI, Switzerland}
\author{R.~Dressler} \affiliation{Paul Scherrer Institut, Villigen PSI, Switzerland}
\author{F.~K\"{a}ppeler} \affiliation{Karlsruhe Institute of Technology (KIT), Institut f\"{u}r Kernphysik, Karlsruhe, Germany}
% ...

\collaboration{The n\_TOF Collaboration (www.cern.ch/ntof)}  \noaffiliation

\date{\today}

\begin{abstract}
We report on the measurement of the $^{7}$Be($n, p$)$^{7}$Li cross section from thermal to approximately 325 keV neutron energy, performed in the high-flux experimental area (EAR2) of the n\_TOF facility at CERN. This reaction plays a key role in the lithium yield of the Big Bang Nucleosynthesis (BBN) for standard cosmology. The only two previous time-of-flight measurements performed on this reaction did not cover the energy window of interest for BBN, and showed a large discrepancy between each other. The measurement was performed with a Si-telescope, and a high-purity sample produced by implantation of a $^{7}$Be ion beam at the ISOLDE facility at CERN. While a significantly higher cross section is found at low-energy, relative to current evaluations, in the region of BBN interest the present results are consistent with the values inferred from the time-reversal $^{7}$Li($p, n$)$^{7}$Be reaction, thus yielding only a relatively minor improvement on the so-called Cosmological Lithium Problem (CLiP). The relevance of these results on the near-threshold neutron production in the p+$^{7}$Li reaction is also discussed.

\begin{description}
\item[PACS numbers]
\pacs{}26.35.+c, 28.20.-v, 27.20.+n


\end{description}

\end{abstract}

\maketitle


Big Bang Nucleosynthesis (BBN) is one of the key elements of the Big Bang Theory, as it describes the production of light elements in the early stage of the Universe \cite{Cyburt2016,Coc2017}. A remarkable agreement is found between BBN predictions and primordial abundances of D and $^{4}$He inferred from observations of objects at high red-shift and/or in ionized hydrogen regions of compact blue galaxies. On the contrary, model predictions seriously overestimate, by more than a factor of three, the primordial abundance of lithium inferred from observation of metal-poor stars (the so-called Spite plateau~\cite{spiteplateau}). Such a discrepancy is known as the Cosmological Lithium Problem (CLiP). Various possible explanations of this problem have been proposed in the last decades, involving the field of Astrophysics, Astronomical observations, non Standard Cosmology and new physics beyond the Standard Model of particle physics, but at present a fully satisfactory solution is still missing.

The primordial $^{7}$Li abundance is dominated by the electron-capture decay $^{7}$Be$ + e^{-}\rightarrow ^{7}$Li$ + \nu_{e}$  \textit{after} the BBN phase of evolution of the early universe. Therefore, a higher rate for neutron or charged-particle induced reactions on $^{7}$Be leads to a lower surviving $^{7}$Be fraction and, ultimately, to a lower $^{7}$Li abundance. To investigate a possible nuclear physics solution to the CLiP, several measurements have been performed in recent years on charged particle reactions responsible for the production and destruction of $^{7}$Be. However, none of these measurements have revealed significant impact on the $^{7}$Be production or destruction  \cite{chargedp}. More recently, an experimental campaign has been undertaken to measure, in some cases for the first time, neutron induced reactions responsible for the $^{7}$Be destruction. This is the case, in particular, for the $^{7}$Be($n, \alpha$) channel, where data in the neutron energy region of interest for BBN were essentially missing. Recently, both a direct measurement \cite{Barbagallo2016} and a time-reversal one \cite{Kawabata2017} have definitely excluded a possible contribution of this reaction to the solution of the CLiP.

The $^{7}$Be($n, p$)$^{7}$Li reaction is responsible for a dominat fraction of the destruction of $^{7}$Be. As a consequence, it plays a key role within BBN models in the determination of primordial lithium. Despite its importance, very few direct measurements exist for this reaction. This cross section was measured in the 80's at the Joint Institute for Nuclear Research (Dubna, Russia) from thermal to 500 eV neutron energy \cite{gledenov} and shortly thereafter at the Los Alamos Neutron Science CEnter (LANSCE) in Los Alamos, USA, from thermal to 13 keV \cite{koehler}.  
Although the former is affected by large statistical uncertainties that make a comparison difficult, the two datasets are somewhat inconsistent with each other, showing a systematic difference of 30\%, on average. Furthermore, the extrapolation to higher energy of the Los Alamos measurement seems to be inconsistent with the cross section determined on the basis of the time-reversal $^{7}$Li($p, n$)$^{7}$Be reaction. 
In summary, at present a consistent and accurate description of the $^{7}$Be($n, p$)$^{7}$Li reaction from thermal neutron energy to the BBN region, i.e. up to several hundred keV, is still missing. Considering the important role of this reaction for the CLiP, new high accuracy data on this reaction, in a wide energy region, would finally clarify the situation. 

The main difficulties of direct measurements are related to the high specific activity of $^{7}$Be (13 GBq/$\mu$g), and to the small Q-value of the reaction (1.64 MeV), leading to the emission of low-energy protons. Both characteristics put severe constraints on the total mass, areal density and purity of the sample, and in turn on the flux of the neutron beam needed for a statistically significant measurement.  In this respect, the recently built high-luminosity experimental area at n\_TOF (EAR2) \cite{Weiss2015} is one of the few time-of-flight installations where such a measurement could be performed. 


The $^{7}$Be($n, p$)$^{7}$Li reaction was studied by combining the capabilities of the two major nuclear physics facilities operating at CERN: ISOLDE and n\_TOF. The sample material was produced at Paul Scherrer Institute (PSI) in Villigen, Switzerland, by radiochemical separation of $^{7}$Be from the SINQ cooling water (details on the sample preparation and characterization are reported in Ref. \cite{Maugeri2018}). The freshly separated material was then implanted on an Al backing at ISOLDE and immediately afterwards irradiated with the pulsed, wide spectrum neutron beam in EAR2 at n\_TOF. Although designed for the production and extraction of radioactive ion beams, the ISOLDE target unit can accommodate imported activity for the efficient preparation of isotopically pure radioactive samples \cite{Borge2017}. 
A $^{7}$Be implanted target of $1.03 \pm 0.3$  GBq total activity, corresponding to $\approx$80 ng in mass was produced, with a purity of about 99\%, the remaining 1\% due to $^{7}$Li contamination. While the implantation was originally designed to produce a uniform sample of 1.5x1.5 cm$^{2}$ area, the obtained deposit was highly inhomogeneous. Two different imaging techniques applied on the sample after the measurement showed that the deposit had a gaussian profile of approximately 0.5 cm FHWM \cite{Maugeri2018}. A correction for the inhomogeneity of the sample had therefore to be applied in the analysis (see below for details).


The main features of the n\_TOF neutron beam in the EAR2 measurement station are the wide neutron energy spectrum, spanning from 2 meV to 100 MeV, the high intensity of $\gtrsim$ 10$^{7}$ neutrons/pulse at the sample position, the low repetition rate, of less than 0.8 Hz, and the good energy resolution (10$^{-3} \leq \Delta E/E \leq $10$^{-2}$ in the energy range of interest for this measurement). More details about the EAR2 neutron beam can be found in Refs. \cite{Weiss2015,Marta2017}.

The experimental setup consisted of a position-sensitive telescope, made of two silicon strip detectors, of 20 $\mu$m and 300 $\mu$m thickness for $\Delta$E and E detection, respectively. Both detectors had a 5x5 cm$^{2}$ active area and 16 strips. The telescope was mounted at a polar angle of 90 degrees, relative to the beam direction, at a distance from the center of the sample of 5 cm. To minimize the energy straggling of emitted protons inside the $^{7}$Be deposit, the sample was tilted relative to the neutron beam direction by 45 degrees. The prompt signal produced by $\gamma$-rays and relativistic particles, the so-called $\gamma$-flash, was used as reference for the determination of the neutron time of flight. Before the $^{7}$Be measurement, the $^{6}$Li($n, t$)$^{4}$He reaction was measured in the same experimental conditions. To this end, a sample of 1.1 mg of $^{6}$LiF was used, with a surface of 1.5 x 1.5 cm$^{2}$, to match the envisaged size of the $^{7}$Be sample. The $^{6}$Li($n, t$) reaction was also used for calibrating the energy deposited in the telescope, by means of the triton peak. More details on the experimental setup can be found in Ref. \cite{Barbagallo2017}. 

The high purity of the sample, the use of a telescope for particle identification, and the very high instantaneous neutron flux of EAR2 resulted in a practically negligible background, in particular the one associated with the natural $\gamma$-ray activity of $^{7}$Be. The only source of background affecting the measurement is related to the $^{14}$N$(n, p)$ reactions in the sample backing. This background was identified and subtracted with a "dummy" sample, i.e. the backing without the $^{7}$Be deposit, and its contribution was found important only for neutron energies above $\sim$500 keV.

\begin{figure}[t]
  \includegraphics[width=9.5cm]{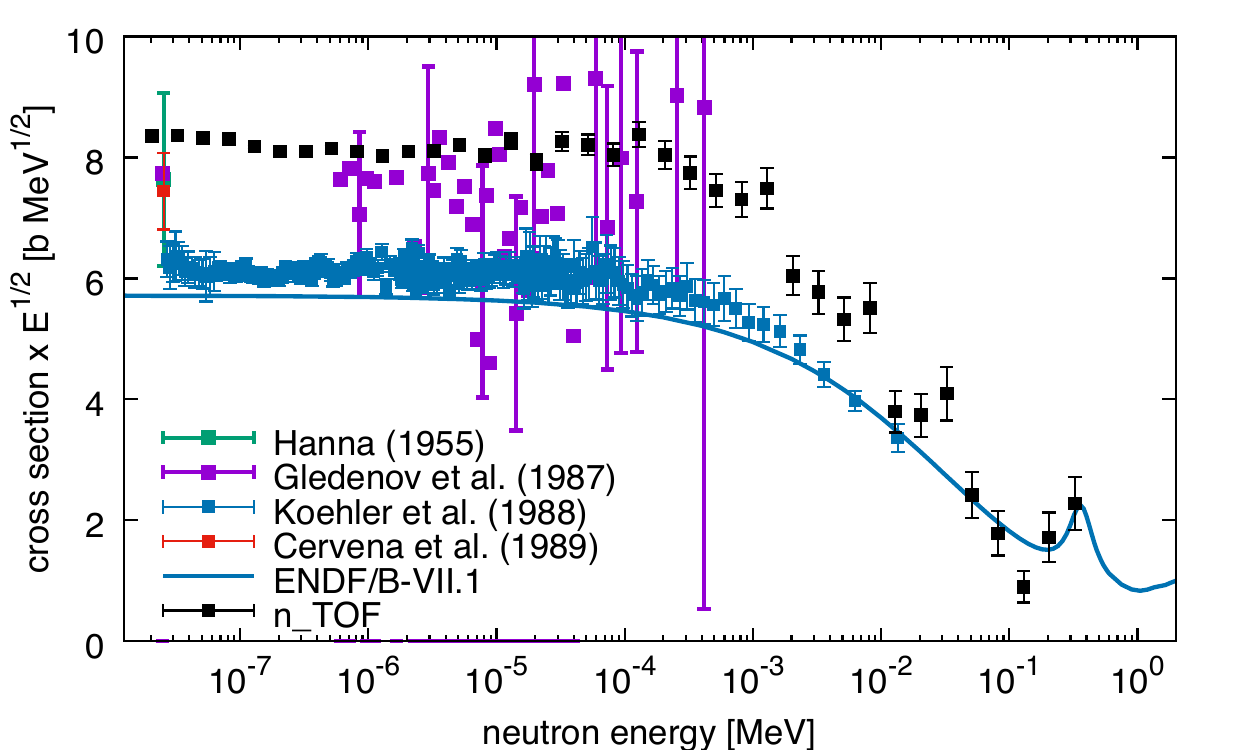}
\caption{The $^{7}$Be($n, p$)$^{7}$Li reduced cross section measured at n\_TOF compared with the results of previous measurements~\cite{Hanna1955,gledenov,koehler,Cervena1989} and with the ENDF/B-VII.1 library~\cite{endfb71}.
\label{fig:fig_xs}}
\end{figure}

\begin{figure}[h]
\includegraphics[width=8.5cm]{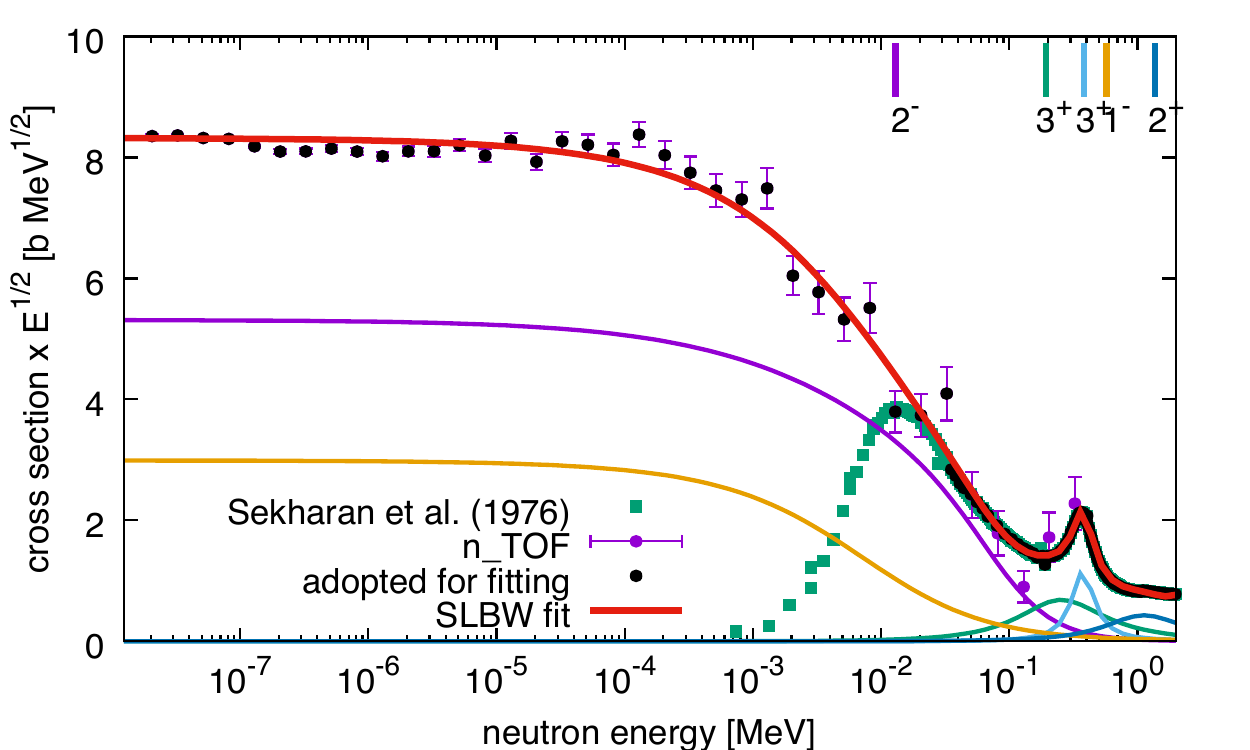}
\caption{Resonance fit of the n\_TOF cross section data, complemented by data from the $(p,n)$ channel (adopted). The contribution from the first five levels (indicated in the upper right corner at their respective excitation energy), are included in the plot (color code online), but all nine states above the neutron separation energy are included in the SLBW fit as described in the text. 
\label{fig:fig_levels}}
\end{figure}

As previously mentioned, the deposit of $^{7}$Be turned out to be highly inhomogeneous and with smaller dimension than originally envisaged, thus being substantially different from the $^{6}$Li sample used as reference. As a consequence, the $^{7}$Be($n, p$) cross section cannot be simply determined from the rate of the $^{6}$Li($n, t$) reaction, considering that the fraction of the neutron beam intercepted by the two samples is significantly different. The cross section can be obtained by

\begin{equation}
\sigma_{n,p}(E_{n}) = \frac{C(E_{n})}{\Phi(E_{n})\times{\epsilon}\times{N_{S}}\times{f_{C}}}.
\label{eq:one}
\end{equation}

Here, C(E$_{n}$) is the number of detected protons (integrated over the whole measurement) in a given neutron energy bin, $\Phi$(E$_{n}$) the total number of neutrons in that energy bin, derived from Ref. \cite{Marta2017}, and $\epsilon$ the detection efficiency, obtained from detailed GEANT4 simulations of the experimental setup. $N_{S}$ is the total number of atoms in the sample and the factor $f_{C}$, introduced to account for the target inhomogeneity mentioned above, represents the convolution of the normalized neutron beam spatial profile and target nuclei distribution and has a dimension of b$^{-1}$. The measured distribution of the sample material \cite{Maugeri2018}, and the neutron profile obtained from GEANT4 simulations of the spallation process and subsequent transport through the vertical beam line \cite{jorge}, were used for the evaluation of $f_{C}$. A consistency check of the method to derive the cross section was carried out by analyzing the $^{6}$Li$(n, t)$ data, with the efficiency and beam-sample convolution factor $f_C$ specifically calculated for the $^{6}$Li sample. The obtained cross section agrees with the standard~\cite{endfb71} within 5\%, in the whole neutron energy range from thermal to 1 MeV. 

The $^{7}$Be$(n,p)$ cross section was extracted relative to that of the $^{6}$Li$(n,t)$ reaction, from the ratio of the number of counts (normalized to the respective total neutron fluence), and taking into account the ratios of the efficiencies and beam-sample convolution factors. This method minimizes the uncertainty, as the energy-dependent flux cancels out, while systematic effects on the simulated efficiencies mostly compensate each other, except at higher energies (see below). Considering that the alignment of the sample relative to the neutron beam is not known to better than a few millimeters, and in view of the difference in the target nuclei distribution of the $^{7}$Be and $^{6}$Li samples, an uncertainty of 8\% was estimated for the ratio of the $f_{C}$'s. Since all other factors in Eq.~\ref{eq:one} are known with a better accuracy, this represents the major source of uncertainty of the present results, up to a neutron energy of 50 keV. 

A final remark concerning the detection efficiency is that, contrary to the $^{6}$Li$(n, t)$ case, for which the known angular distribution above 1 keV was used in the simulations, the emission of protons in the n+$^{7}$Be reaction was assumed to be isotropic at all energies. This assumption was verified "a posteriori", on the basis of the levels of $^{8}$Be contributing to the cross section. In fact, as will be shown later on, up to 50 keV the cross section is dominated by negative parity compound levels, favoring s-wave proton emissions. Above this energy, this assumption may not hold anymore, leading to an additional 10\% systematic uncertainty in the extracted cross section. 

Figure \ref{fig:fig_xs} shows the background-subtracted reduced cross section (i.e. the cross section multiplied by the square-root of the neutron energy) of the $^{7}$Be($n, p$)$^{7}$Li reaction, as a function of neutron energy, compared with the two previous direct measurements and with the current ENDF evaluation. 
In the figure only the statistical errors are shown. The systematic uncertainty, mainly related to the sample inhomogeneity, is 10\% from thermal to 50 keV, and could reach 15\% above it, due to the estimated effect of the angular distribution assumption.
The present data are 35\% and 40\% higher than the ones of Koehler \textit{et al.} ~\cite{koehler} and of the ENDF/B-VII.1 evaluation~\cite{endfb71}, respectively, while they are consistent with the results of Hanna~\cite{Hanna1955}, Gledenov \textit{et al.}~\cite{gledenov}, and \v{C}erven{\'a} \textit{et al.}~\cite{Cervena1989} at thermal neutron energy. Our experimental value sets at  $ 52.3 \pm 5.2$  kb. 

Even though the n\_TOF measurement covers a wide energy range up to $\sim$ 325 keV, in order to derive a rate on the $^{7}$Be$(n,p)^{7}$Li reaction at thermal energies of interest for Big-Bang Nucleosynthesis calculations, the cross section of the present measurement has been complemented by data derived from the time reversal reaction, which provides accurate information in the upper energy side of the energy spectrum, from 35 keV to $\approx$ 2 MeV. To this end, the data of the $^{7}$Li$(p,n)^{7}$Be cross section from Sekharan \textit{et al.} \cite{Sekharan1976} have been used in conjunction with the specific detailed-balance relation:

\begin{equation}
\sigma_{n,p} = \frac{k_{p}^2}{k_{n}^2} \sigma_{p,n} 
\label{eq:two}
\end{equation}

where $k_{p}$ and $k_{n}$  are the incident proton and outgoing neutron relative momenta in the center-of-mass systems. The $R$-matrix single-level Breit-Wigner (SLBW) formalism, which is appropriate for the present situation, has been used for fitting the cross section in the full energy range, including all nine states in $^{8}$Be above 18.899 MeV (the neutron separation energy) and up to 22 MeV. The excitation energy of each state, as compiled in the ENSDF library \cite{ENSDF}, has been kept constant in the fit, while the neutron and/or the proton widths have been allowed to vary, starting from the values reported in the library. The final result is shown in Figure \ref{fig:fig_levels} by the red curve. 

\begin{figure}[h]
\includegraphics[width=8.5cm]{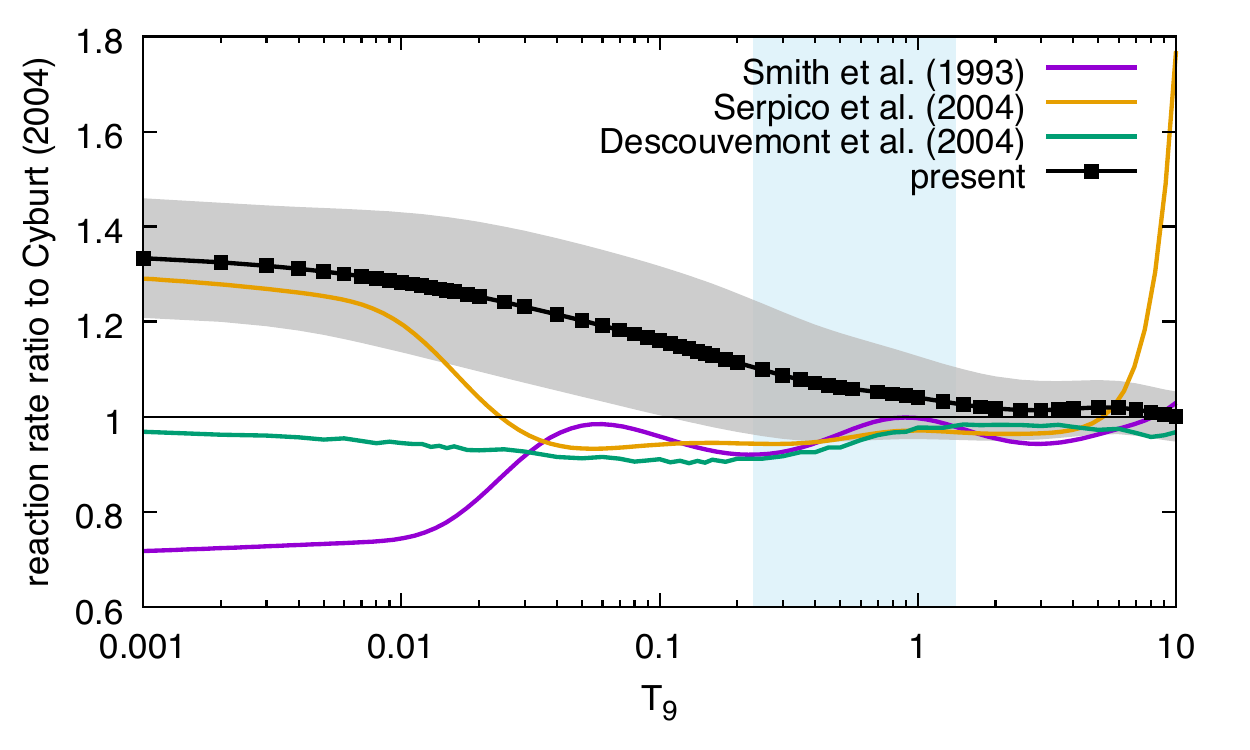}
\caption{Rates of the $^{7}$Be$(n,p)^{7}$Li reaction relative to Cyburt~\cite{Cyburt2004}. The present result is shown with the associated error band and the rates of Smith {\it et al.}~\cite{Smith1993}, Serpico et al.~\cite{Serpico2004}, and Descouvemont et al.~\cite{Descouvemont2004} are shown for comparison. The temperature range of BBN with larger impact on the lithium yield is indicated by the vertical band.}
\label{fig:fig_rr}
\end{figure}

From the fitted data, the cross section averaged over the energy distribution corresponding to the temperature of interest in standard BBN model calculations has been derived in the temperature range, $0.001 \leq T_{9} \leq 10$, where T$_{9}$ indicates the temperature in units of $10^9$ K. The resulting reaction rate can be accurately described by an analytical expression (see the supplementary material \cite{twikipage} for all details)
\begin{equation}
\begin{split}
N_{A}  <\sigma v> = a_{0} (1 + a_{1}T_{9}^{1/2} + a_{2} T_{9} + a_{3}T_{9}^{3/2}  + \\ 
+ a_{4}T_{9}^{2} + a_{5}T_{9}^{5/2}) + a_{6}(\frac{1}{1+13.076 T_{9}})^{3/2} + \\ 
 + a_{7}T_{9}^{-3/2} e^{-b_{0}/T_{9}} 
\end{split}
\label{nsv}
\end{equation}
in units of  cm$^{3}$/s/mole when $a_0=6.805\times10^{9}$, $a_1=-1.971$, $a_2 = 2.042$, $a_3 = -1.069$, $a_4=0.271$, $a_5=-0.027$, $a_6=1.961\times10^{8}$, $a_7=2.890\times10^{7}$ and $b_0=0.281$.

A comparison of the present reaction rate with other rates commonly adopted in BBN calculations is shown in Figure \ref{fig:fig_rr}. It can be seen that the present rate is significantly higher in a wide range up to $T_{9} \approx 1$. 

We have performed standard BBN calculations using an updated version of the AlterBBN code~\cite{Arbey2012}, adopting a neutron mean life-time of $\tau_{n} = 880.2 $ s and $N_{\nu} = 3$ neutrino species. The remaining additional parameter of the adopted standard cosmology is the baryon-to-photon number density ratio, $\eta_{10}$  ($\eta$ in units of $10^{-10}$), determined either from the CMB observation ($\eta_{10} = 6.09 \pm 0.06$) or as a range considered to be allowed by other light nuclei observables ($5.8 \leq \eta_{10} \leq 6.6$)~\cite{PDG2017}. We have adopted an updated set of reaction rates~\cite{starlib} for the twelve most important reactions (as defined in~\cite{Cyburt2004}) and details on each individual rate are provided in the supplementary material~\cite{twikipage}. The results for the Li/H production are shown in Table  \ref{table}. The uncertainty associated to the Li/H yield in the present calculation is of the order of 8\%, evaluated adopting the upper and lower limits of the rate, as shown in Figure~\ref{fig:fig_rr}. The new results reported in this work essentially leads to, at most, a 12\% decrease in the Lithium production relative to previous estimations, a change that does not have a significative impact on the cosmological lithium problem.

\begin{table}[t]
\begin{ruledtabular}
\centering
\begin{tabular}{lcl}
         $^{7}$Be$(n,p)^{7}$Li rate                                                     &  $\eta_{10}$          & Li/H yield           \\ \hline 
Cyburt (2004)~\cite{Cyburt2004}         &      6.09                & $5.46$              \\
This work (Eq.~\ref{nsv})                                              &      6.09         & $5.26 \pm 0.40 $          \\
                                         &  $ 5.8 - 6.6 $     & $ 4.73 - 6.23 $          \\
Observations~\cite{Cyburt2016}      &       & $1.6 \pm 0.3$    \\
\end{tabular}
\caption{BBN $^{7}$Li/H abundance (in units of $10^{-10}$ and for different $\eta_{10}$ values~\cite{PDG2017}), obtained with the rate determined in this work for the $^{7}$Be$(n,p)^{7}$Li reaction (see text for all other reaction rates). The Li/H abundance calculated with the previously adopted rate of reference~\cite{Cyburt2004} is also reported, for comparison.} 
\label{table}
\end{ruledtabular}
\end{table}

\begin{figure}[h]
  \includegraphics[width=9.cm]{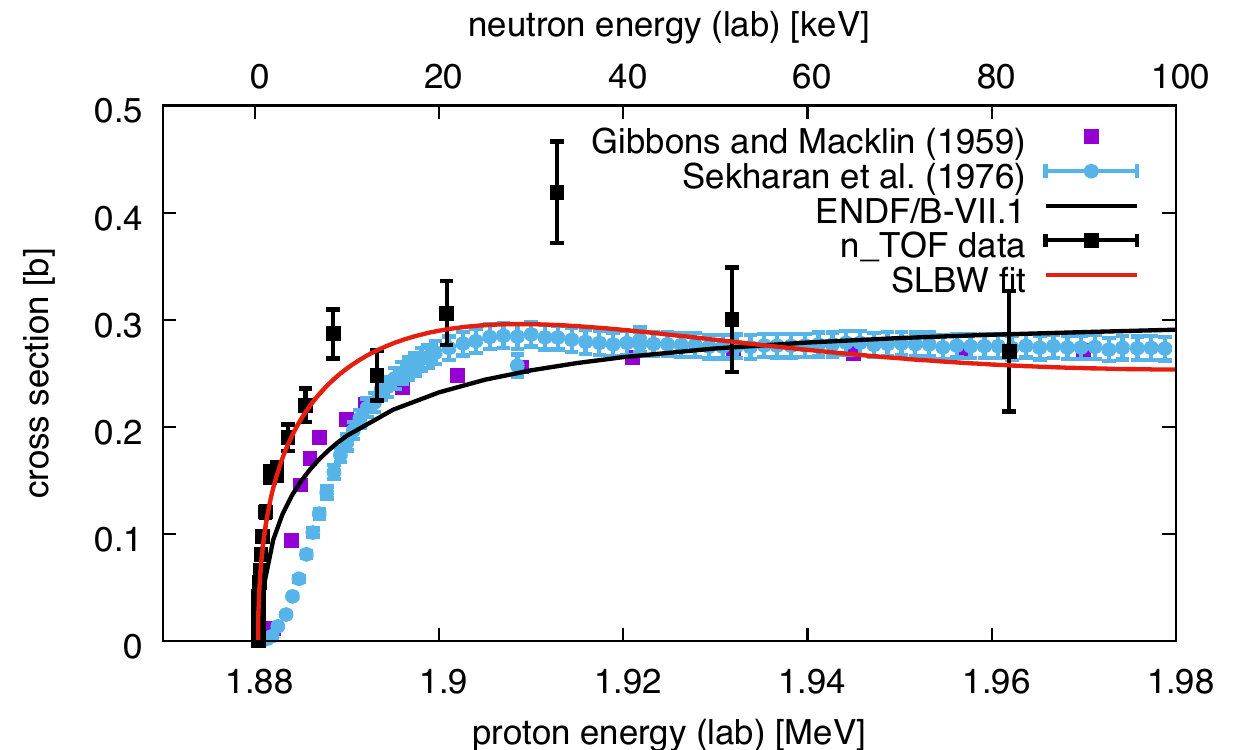} 
\caption{The cross section of the $^{7}$Li($p, n$)$^{7}$Be (black symbols), obtained by time-reversing the n\_TOF data of the $^{7}$Be($n, p$) reaction. For comparison, the data from measurements of the near-threshold $(p,n)$ reaction (Refs.~\cite{Gibbons1959,Sekharan1976} ) are also shown.  \label{fig:fig_7li}}
\end{figure}

The present data can also provide information on the cross section of the $^{7}$Li($p, n$)$^{7}$Be reaction, one of the most important reactions for neutron production at low-energy accelerators, widely used for a variety of applications. In particular, for proton energies slightly above the reaction threshold, namely around 1912 keV, the forward-emitted neutrons from thick $^{7}$Li targets show a quasi-Maxwellian energy distribution that mimics the stellar neutron spectrum at kT$\sim$25 keV, a feature that makes this neutron source very attractive for astrophysics-related studies \cite{ppnp}. The excitation function of this reaction is of crucial importance for an accurate estimate of the neutron yield and spectrum in thick targets (see for example Ref. \cite{lee,herrera}).
Direct measurements of the $^{7}$Li($p, n$)$^{7}$Be reaction cross section near the threshold (E$_{p}$=1880.3 keV) are difficult to perform, due to the low energy of the emitted neutrons ($\sim$30 keV at the threshold), the need of a proton beam of stable and well calibrated energy, and the relatively poor resolution related to the energy loss of protons inside the target. For these reasons, discrepancies exist between various datasets near threshold, as shown in Refs. \cite{friedman,guido}. On the contrary, the $^{7}$Be($n, p$) data are not affected by those experimental problems, and can therefore be used for a very accurate determination of the excitation function through Eq. \ref{eq:two}. The results are shown in Figure \ref{fig:fig_7li}. Compared with direct measurements, the extracted excitation function shows a much faster rise above the threshold, as they are not affected by the resolution problems of direct measurements. The new data can be used for more accurate calculations of the reaction yield and neutron spectrum in the near-threshold $^{7}$Li($p, n$) reactions.

In conclusion, a new measurement of the $^{7}$Be($n, p$)$^{7}$Li reaction from thermal to $\sim$ 325 keV neutron energy has been performed at n\_TOF with a high purity sample produced at ISOLDE, demonstrating the feasibility of neutron measurements on samples produced at radioactive beam facilities. The cross section is higher than previously recognized at low energy, by $\sim$40\%, but consistent with current evaluations above 50 keV. The new estimate of the $^{7}$Be destruction rate based on the new results yields a decrease of the predicted cosmological Lithium abundance of $\sim$10\%, insufficient to provide a viable solution to the Cosmological Lithium Problem. The two n\_TOF measurements of $(n,\alpha)$ and $(n,p)$  cross sections of $^7$Be can finally rule out neutron-induced reactions, and possibly Nuclear Physics, as a potential explanation of the CLiP, leaving all alternative physics and astronomicalß scenarios still open. 

\begin{acknowledgments}
The authors wish to thank the PSI crew A. Hagel, D. Viol, R. Erne, K. Geissmann, O. Morath, F. Heinrich and B. Blau for the $^{7}$Be collection at the SINQ cooling system. This research was partially funded by the European Atomic Energy Community (Euratom) Seventh Framework Programme FP7/2007-2011 under the Project CHANDA (Grant No. 605203). We
acknowledge the support by the Narodowe Centrum Nauki (NCN), under the grant UMO-2016/22/M/ST2/00183.
\end{acknowledgments}

\bibliography{pap}

\end{document}